\documentclass[smallextended]{svjour3}

\usepackage{latexsym}
\usepackage{amsmath}
\usepackage{amsfonts}
\usepackage{amssymb}
\usepackage{graphicx}
\usepackage{setspace}

\newcommand{\fE}{\mathcal{E}}
\newcommand{\mf}{\leftrightarrow}
\newcommand{\prs}[1]{{\left(#1\right)}}

\newcommand{\chs}[1]{{\left\{#1\right\}}}
\newcommand{\prob}[1]{{\mathcal{P}\prs{#1}}}
\newcommand{\probi}[2]{{\mathcal{P}_{#1}\prs{#2}}}

\newcommand{\cut}[1]{}

\begin{document}

\title{Falsifiability of Isolated Spacetime Regions}
\author{Roberto C. Alamino}
\institute{Non-linearity and Complexity Research Group, Aston University, Birmingham B4 7ET, UK}

\maketitle

\begin{abstract}
In this work it is pointed out that some physical theories, even being themselves falsifiable, predict the existence of regions of spacetime which are not falsifiable with relation to each other due to their impossibility of mutually exchanging information as, for instance, before and after the event horizon of black holes. If we require scientific theories to be falsifiable, an isolated region should be discarded from scientific models developed by observers in other regions. Here it is proposed that their existence can satisfy a weaker falsifiability condition, here called conditional asymptotic provability, which extend scientific reasoning through Bayesian inference. Limitations and some epistemic consequences of this proposal are discussed. 
\end{abstract}


\section{Introduction}
\label{section:Intro}

The existence of alternate realities or parallel universes in addition to our own is a classic problem both in physics and philosophy. It can be traced back to the dialogues of Plato\cite{Plato} in which he stated that all possibilities should be actualised in the universe, what later became known as the \textit{principle of plenitude} and was restated with variations by several different philosophers. For instance, \textit{modal realism} \cite{Lewis86} suggests that every possible \textit{world} -- another term used to denote these alternate realities -- exists independently of any requirements in terms of physical laws. Such a proposal, however, exchanges the question of `existence' by that of `actuality' without clearly defining the two terms. Although modal realism claims that all worlds somehow exist, they are actual only relative to other entities belonging to the same world, which are called \textit{worldmates}. This proposal literally suggests that all Shakespeare's characters are actual in their own worlds, but the meaning of `actual' remains elusive. 

If one considers modal realism and similar proposals as \textit{hypotheses} about the natural world, there are some important formal issues that need to be addressed. In particular, it is of fundamental importance to find a precise meaning for the term `possible worlds' and characterise more precisely the meaning of `to exist' and 'to be actual'. At the bare minimum, the limitations of these concepts relative to physical theories should be understood. These are not trivial questions. For instance, the original version of modal realism suggests that possible worlds should be considered to be generated by a recombination principle which, however, becomes equivalent to restating that everything is possible. Attempts to constrain the possibilities become too artificial in the proposed framework as they are based on the exclusion of certain sets of worlds from it, defeating the spirit of the original idea. Philosophically, this issue seems far from being settled \cite{Russell15}.

The problem is particularly relevant in modern physics, where parallel universes appear in many different guises \cite{Tegmark05}. The ensemble formed by all these universes within a certain framework is known as the \textit{multiverse}, although the way this term is used may vary widely \cite{Carr07}. But the idea of parallel universes in physics precedes these relatively recent studies, being for instance the foundation of Everett's \textit{many-worlds interpretation} of quantum mechanics \cite{Everett57,Bradley11}. Although older than modal realism, the many-worlds interpretation suggests a similar scenario, but restricts the set of possible worlds to histories of systems which forcibly obey the basic laws of quantum mechanics.
 
A more ambitious idea proposed by Tegmark \cite{Tegmark98,Tegmark08} and named the \textit{mathematical universe} is a version of modal realism which is less restrictive than many-worlds. It suggests that every mathematical structure can be considered to have physical existence and, therefore, all universes modelled after these structures also do. Given its close relation to modal realism, we call this proposal \textit{mathematical realism}. It is argued that mathematical realism trivially explains Wigner's famous observation about the ``unreasonable effectiveness of mathematics in the natural sciences'' \cite{Wigner60}. According to it, there is no mystery in the fact that the universe can be described by mathematical models with unbelievable precision because every universe that can be described this way is ultimately bound to exist. The possible existence of non-mathematical universes seems to be outside the scope of the proposal.  

Both many worlds and mathematical realism suffer from the same problem of defining the meaning of `to exist'. The concept of `absolute existence' has yet to be formulated in a totally satisfactory and uncontroversial way, not only in physics but also in philosophy \cite{McGinn00,Ross13}. This work has no intention of contributing to that particular line of research. In order to make progress in our intended direction, we take the pragmatic position that, although there is still no such definition available, the concept of absolute existence has a meaning which coincides, when appropriate, with our physical intuition about it. 

Another topic to which we shall not contribute is choosing whether either modal or mathematical realism is better. We assume that it is an experimental fact that mathematical models used by physics can approximate reality and restrict our analysis to them from here on. Combining our two aforementioned assumptions, it should be possible in principle to define a binary operator in every physical model which acts on every object $O$ of the theory (particle, field, system etc.) in such a way that $E(O)=1$ means that $O$ exists (in the absolute sense of the previous paragraph) as a physical entity and $E(O)=0$ that it does not. An actual proof of this possibility is elusive though.

The concept of existence is closely related to the question of whether or not a physical theory is a correct description of nature. Although mathematical realism would imply that all mathematical theories are correct because the mathematical models represented by them do exist, we know by experience that not all of them describe \textit{our} reality. Some consistent mathematical descriptions will give wrong answers to the results of experiments and therefore should be discarded. If one could prove that a physical theory is correct beyond doubt, there would be a chance that its component objects indeed exist in an absolute sense.  

There are two problems with this reasoning. First, the number of possible hypotheses about the physical world is too large. A more realistic objective would be to identify a subset formed by those which have a possibility of being validated. This is known as the \textit{delimitation problem} and has a key role in separating science from superstition. One of the difficulties faced by this task is that, to the extent of which time has no future boundary, it is impossible to prove that a theory is right. No matter the amount of supporting evidence accumulated by a theory up to some point, there is no possible proof that it will work even in the following instant. One solution for the delimitation problem was proposed by Popper in 1934 \cite{Popper34} and relies on a very precise property of hypotheses called \textit{falsifiability}. Whether or not this is a definitive solution is still a source of controversy \cite{Dawid18}. 

A hypothesis is falsifiable if there is an experiment (which is realisable in principle, but might not be in practice) for which one of its possible results will rule out the hypothesis with certainty. If this condition is not satisfied, it means that the hypothesis has no actual prediction about \textit{any} conceivable experiment. In other words, every measurement is trivially consistent with the hypothesis -- the truth value of the hypothesis has no actual consequences in the physical world (although it might have psychological ones). 

The above line of reasoning led Popper to argue that falsifiability should be the main property of scientific theories. If absolute existence as mentioned before can be actually defined, models which are not falsifiable cannot make non-trivial predictions, having no observational consequences even if the reality they describe exists. Intuitively, in order for a hypothesis to be falsifiable, a good degree of precision in its formulation becomes necessary. Scientific models (in particular, from physics) are hypotheses usually built on rigorous mathematical structures and therefore more probable to be falsifiable. Still, not all scientific hypotheses being currently studied are \textit{known} to be falsifiable (for instance, string theory \cite{Woit11}). There is nothing wrong with that, but once the theory is \textit{proved} to be not falsifiable, common sense suggests it should be better to discard it from our attempt to describe nature rationally. 

Let us now state the main problem identified in this work: if we assume that the concept of absolute existence is meaningful, it is possible that different regions of spacetime which are not falsifiable with respect to each other exist. However if two regions $M_1$ and $M_2$ exist, but the existence of $M_2$ cannot be falsified by $M_1$ (or vice-versa), science constrained by falsifiability cannot assess it. In fact, the existence of $M_2$ from the point of view of $M_1$ becomes pure philosophical speculation. We reach the very awkward situation in which, although both regions do exist, it is not rational for observers in $M_1$ to say that $M_2$ does. We call it the \textit{isolated regions problem}.

The above question has been raised in the context of multiverse theories before, but in a different way. While most discussions focus on whether or not the models are valid as scientific hypothesis \cite{Carr08}, we start by assuming the existence of isolated regions and propose a (partial) solution for the problem. We will return to this point in more details later on.

Because falsifiability relies on the possibility of an observer measuring the results of experiments, it requires the exchange of information between the observer and the experiment. This means that information should be able to flow from the spacetime location of the experimental set up to that of the observer. We identify here some physical models in which this indeed happens -- there are predicted regions which cannot exchange information between themselves.

Here, we propose a partial solution to this problem -- to include in the scope of science a different class of theories which cannot be proven to be wrong, but whose \textit{probability} of being wrong can be continually assessed and re-evaluated. In this way, although the existence of $M_2$ to $M_1$ still cannot be falsified, one can make some quantitative objective assessment of the probability that its existence is true.

The restrictive nature of falsifiability is due to the fact that it is binary concept -- a hypothesis \textit{is} or \textit{is not} falsifiable, two mutually exclusive possibilities. Our proposal introduces an alternative point of view -- that if we accept the limitation of not knowing for certain whether a theory is also wrong, a probabilistic relaxation of this condition allows the definition of a concept we call conditional asymptotic provability (CAP). The truth value of hypotheses satisfying CAP can then be continuously assessed in face of experimental evidence through Bayesian inference methods.

Under certain conditions, we will show that isolated regions of spacetime might obey CAP implying that, by using Bayesian inference one can gradually increase the confidence in their existence from another region. Under this framework, hypotheses which are not falsifiable obviously remain so, but one can nevertheless asses their level of confidence  adding a new epistemic level to the delimitation problem and extending the scope of scientific inquiry. 

The rest of this paper, which is intended to formally specify the general problem and its proposed resolution, is structured as follows. In section \ref{section:NFR} we give a more detailed definition of what is an isolated spacetime region. Section \ref{section:fals} formalise the concept of falsifiability in probabilistic terms. Two conditions are then proposed to classify hypotheses as falsifiable and the consequences of them are worked out. Then, in section \ref{section:CAP} we introduce the idea of conditional asymptotic provability and analyse the conditions under which it can occur. The final section of this work is dedicated to a summary of the results and additional discussions.

\section{Isolated Regions}
\label{section:NFR}

Consider two existing physical entities $A$ and $B$, i.e. $E(A)=E(B)=1$, for which it is impossible to exchange information either directly or indirectly, where information is to be understood in the sense of Shannon \cite{Cover91}. Physically, information can be exchanged if there is a possibility of interaction between two entities, mediated or not by a third one. For instance, interactions between elementary fermions are always mediated by bosons.

When even indirect interaction is not possible, one entity cannot falsify the existence of the other. In this case, either $A$ is not falsifiable to $B$ (but $B$ is to $A$), $B$ is not falsifiable to $A$ (but $A$ is to $B$) or both are mutually not falsifiable. Notice that the possibility that information can flow in only one way, for instance, from $A$ to $B$ but not from $B$ to $A$ is left open. If $A$ is falsifiable to $B$, but $B$ is not to $A$, we write $A\rightarrow B$ or equivalently $B\leftarrow A$. This notation intends to convey the idea that information can flow in the direction of the arrows. When information can flow both ways, we say that the entities are \textit{mutually falsifiable} and write $A\mf B$. Entities are clearly falsifiable to themselves, i.e., $A\mf A$ for every entity such that $E(A)=1$. If information cannot flow either way and entities are not falsifiable to each other, we write $A\parallel B$ and say that they are \textit{isolated} from each other.

\begin{figure}
	\centering
	\includegraphics[width=5cm]{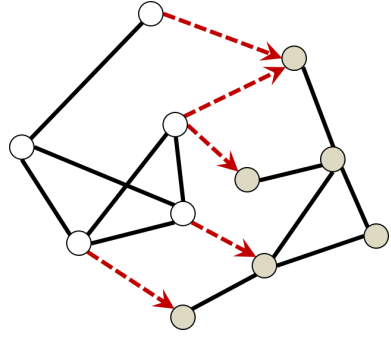}
	\caption{Representation of the falsifiability relations between entities as a directed graph. Directed (red) edges represent the direction of information flow while non-directed (black) edges represent flow in both directions.} 
	\label{figure:regions}
\end{figure}

One can represent the above relations between entities in a given patch of spacetime by a directed graph in which each entity is represented by a vertex. An arrow from vertex $A$ to vertex $B$ indicates that information can flow from $A$ to $B$, but not in the opposite direction. When information flows both ways, we indicate it by an undirected edge for simplicity. Fig. \ref{figure:regions} shows an example of this kind of graph.

Notice the presence of a boundary between two sets of vertices -- hollow and full -- formed by all the directed edges. We will call this boundary an \textit{information horizon}. Information can flow from any vertex to the full vertices, but cannot flow from full vertices to hollow ones. Inside each set of vertices, existence is falsifiable. This is not true in general for the different sets. Sets of vertices which are mutually not falsifiable would correspond to disconnected pieces of a graph. One can now recognise fig. \ref{figure:regions} as a discretised toy model for a neighbourhood of a classical Schwarzschild black hole. The information horizon represents the event horizon, the hollow vertices the exterior of the black hole and the full vertices its interior.

We now claim that falsifiability only makes sense in the relative way defined above. Suppose we could say that an entity $A$ is falsifiable in an absolute sense. The only non-relative meaning for this is if $A$ is falsifiable to every other existing entity. Therefore, if there is one entity that cannot falsify $A$, absolute falsifiability makes no sense. But the same is true for every other existing entity, implying that every pair of entities are mutually falsifiable. This means that absolute falsifiability only makes sense if every entity is falsifiable to every other entity. However, we will soon show examples where this is not true. Therefore, absolute falsifiability cannot exist. 

Because all physical entities will exist in a certain spacetime region, and even spacetime itself can be seen as a physical entity, we can actually associate regions of spacetime to sets of entities. Let us consider two such regions $M_1$ and $M_2$. We say that $M_1\rightarrow M_2$ if there is at least one entity in $M_1$ which is falsifiable to at least one entity in $M_2$. Because all entities inside a certain region are mutually falsifiable, the flow information implies that, every entity in $M_1$ is falsifiable by every entity in $M_2$ therefore. This allows us to unambiguously define the relation $M_1\rightarrow M_2$, i.e., region $M_1$ is falsifiable to region $M_2$. 

We can now justify our use of exchange of (Shannon) information to define falsifiability. The objective is to avoid any ambiguities originating from entities connected by entanglement which, although representing non-local correlations, will not allow any information to be transmitted unless it is mediated locally by one of the four known physical interactions \cite{Ghirardi80}. If two entangled particles somehow end up finding themselves in two regions which are not mutually falsifiable, one can still be considered to be non-falsifiable to the other. Whether or not the existence of non-local quantum correlations would make falsification possible is not clear at the moment and will not be further discussed here.

Each separate world from modal realism, can be associated to an isolated region and the concept of worldmates to that of mutually falsifiability. The same is trur for the the parallel universes in Everett's many-worlds interpretation and the realities in Tegmark's mathematical realism. But these are not the only examples one can find. 

Some actual physical models inspired by string theories propose that our observable universe is a lower dimensional object, called a ``brane'', embedded in a higher dimensional structure that can be occupied by other similar objects (e.g., brane-world scenario \cite{Randall99} or ekpyrotic universe\cite{Khoury01}). In all these models, however, branes are obliged to interact via gravity and can therefore exchange information, not corresponding to the kind of isolation we are concerned with. The possibility of two isolated branes which do not interact at all cannot be however discarded in principle.

Eternal inflation models \cite{Guth00} predict the existence of \textit{pocket universes}, separate regions of spacetime originated from the inflation of a false vacuum which had never interacted with our universe and, in some cases, are predicted to never interact. Another example is the \textit{baby universes} originated from classical black hole solutions \cite{Hsu07}, in which the topology of spacetime changes as a region detaches itself completely from the original manifold and might never be reunited with it again. 

Whether or not the above models are falsifiable themselves is not known. A more striking scenario comes from one of our most well-tested physical theories -- General Relativity (GR). The mathematical structure of GR is consistent and it is generally uncontroversial that it is a falsifiable theory in the sense that it makes falsifiable predictions. In fact, so many of these predictions have been confirmed that our belief that GR is correct as a good approximation for our world given our experimental precision is extremely high.

However, using the framework we developed above, we now show that there are regions of spacetime predicted to exist by GR which are not falsifiable relative to others. One example is the Schwarzschild black hole as suggested above. As mentioned, as far as we know, an observer located behind the event horizon can receive information from the outside, but no information can go back once it has crossed the horizon \cite{Chandrasekhar83}.   

This situation divides the spacetime in the regions before and after the horizon, which we call respectively $O$ and $I$ with $O\rightarrow I$ -- although $O$ is falsifiable to $I$, $O$ is not falsifiable to $I$. Entities in $I$ can record the results of experiments in $O$ by receiving, for instance, light signals from it, but entities in $O$ cannot do any measurements of entities in $I$. The unavoidable conclusion is that, whatever predictions GR does about $I$, they are not falsifiable.   
  
One possible argument against the above observation is that, if the black hole has its origins in the collapse of a star, the region that becomes gradually engulfed by the horizon existed before and there is no evidence that the formation of the horizon would destroy it. The argument works for any causal horizon. Because the receding speeds due to the expansion of the universe are not constrained by the speed of light, two regions in causal contact at the present might become isolated from each other in the future. While previous contact might be evidence for mutual existence, two regions that start their existence already at points that are receding at speeds exceeding that of the light from each other will not be able to exchange information at any time in the future. In this case an additional argument of symmetry could be raised. If there is nothing special about any of the regions relative to the other, why should one stop existing and the other not?

On the other hand, although one might argue that the branching or merging of regions would be enough to assert their possible existence after they become isolated, that is not correct. Whatever processes separate the regions, from the point of view of the other they still need to be falsifiable. That is because the number of hypotheses that can be raised to explain the existence or lack of existence of the other region is infinite. In the example of the expansion of the universe, one can always postulate that a random quantum fluctuation will destroy one of the other regions and those asking the questions are the lucky remaining ones, a version of the anthropic principle.  

Likewise, for the case of a black hole, an alternative theory is that the horizon simply works as a limit to the spacetime and everything that ``crosses'' it stops existing. Alternatively, one could sew beyond the horizon any kind of parallel universe. That would not affect GR outside the black hole as it would not have any measurable effects. Something similar has been proposed under the name of \textit{firewall hypothesis} concerning what happens with an observer when passing through the horizon of a black hole. While classically it seems that nothing would happen, quantum effects put in doubt the survival of a diving observer, which might be incinerated during the crossing \cite{Almheiri13}, and actually the very existence of spacetime beyond it. Susskind have strongly argued against it \cite{Susskind16}, but under the strict umbrella of falsifiability, the discussion has only philosophical value.

But Susskind's position, shared by several others, resonates with many of us for a series of reasons. All of them comes from our intuition born from experience with other physical theories where extreme events like a piece of spacetime being erased do not happen. Most of our theories rely on smooth changes at borders, on symmetry arguments and so on. Still, all of them have no fundamental precedence over observation and, therefore, do not preclude falsifiability. The fact that such characteristics have been observed to be true time and over again cannot guarantee that they will always be true, but each time they happen we do have the right to be more confident about it. Why we have this right is a question of probabilistic inference, more precisely Bayesian inference. In this way, our suggestion is natural -- let us use this right and, aware of its limitations, include probabilistic inference in our epistemological arsenal to explore the physical world.

\section{Falsifiability}
\label{section:fals}

Because our argument is based on Bayesian inference, we need to know how to express falsifiability in these terms. Consider a certain hypothesis $H$ whose truth we want to evaluate in face of some acquired data $D$. More than often, this evaluation has to be probabilistic because not every dataset $D$ will be enough to decide without any doubt. In the Bayesian framework, we say that we want to evaluate the \textit{posterior} probability of $H$ being true after we acquired the dataset $D$. For this, we use the shorthand conditional notation $\prob{H|D}$ (probability of $H$ given $D$). Because this is a conditional probability, we can use Bayes' theorem to write
\begin{equation}
  \prob{H|D} = \frac{\prob{D|H}\prob{H}}{\sum_H \prob{D|H}\prob{H}},
\end{equation}
where the sum over $H\in\chs{F,T}$, $F$ meaning false and $T$ meaning true, is just a normalisation constant and can be omitted whenever only the relative value of $H$ being true or false is important.

This is usually how Bayesian inference is justified in general. The probability $\prob{H}$ is called the \textit{prior} probability of $H$ and is supposed to encode all previous information that would allow us to make an educated guess about the validity of $H$ before including $D$, creating a bias on the $H$ truth value. Bayesian inference can revert the bias in case new data provides enough evidence for it. If we cannot guess any result beforehand, we simply start with equal odds to both possibilities. The probability $\prob{D|H}$ is called the \textit{likelihood} of $H$ given $D$ and encodes a model of the measurement that needs to accompany the hypothesis. Without this model, there is no way to connect the dataset to it. The confuse nomenclature of the likelihood, which one would expect to be of $D$ given $H$, is a quirk from its origins in statistics. That is why, most of the time, we say only \textit{likelihood} or simply the probability of $D$ given $H$. In this way, we can update our prior estimate on the validity of $H$ after acquiring $D$. Except that, at this point, this is not exactly true.     

The reason is that Bayes' theorem comes straight from the definition of conditional probabilities and has no time sense embedded in it, while Bayesian inference does. If one wants to use it as an update equation, it is necessary to prove that the probability on the left hand side refers to a time \textit{after} the one in which we estimated $\prob{H}$. This very subtle point is well discussed in \cite{Caticha11}. Luckily, the final result is formally the same and we can forget about these details most of the time.

There is strong evidence that the formal framework of science should include Bayesian inference \cite{Bovens03,Jaynes03,Howson06,Iranzo08} as long as its limitations are taken into account \cite{Dawid13,Dawid18}, although there is still some discussion about the details of how this can be accomplished \cite{Henderson13}. It can be shown that Bayes' rule itself is a consequence of the more fundamental Principle of Maximum Entropy \cite{Caticha11}, a generalisation of Laplace's Principle of Insufficient Reason, which is nothing but the statement that the objective of rational inference is to extract information from a dataset with the least degree of bias. 

Within the Bayesian framework, probabilities always encode the current state of knowledge about some hypothesis once all the available data (until that moment) have been taken into consideration. Whether or not the proposition is falsifiable does not preclude this attribution. For instance, solipsism, the doctrine that only your mind exists and nothing else, is fundamentally not falsifiable. Still, there is nothing that forbids attributing a probability for it being true or false. Another hypothesis which is not strictly falsifiable is fine-tuning, which also can be analysed using Bayesian/probabilistic methods \cite{Wells19}. The possibility of attaching probabilities to hypothesis which are not falsifiable is nothing magical, only a consequence of the fact that Bayesian inference encodes formal rules of reasoning for general hypotheses. But exactly because of that, it can be seen as a consistent mathematical framework which can be used to extend the limits of rational thought and allow for rational mathematical consideration of hypotheses which are usually in the realm of philosophy but not of science.

Let us see then, how falsifiability can be expressed in probabilistic terms and what are the consequences of using Bayesian inference together with it. Suppose, once again, that we have a certain hypotheses $H$ and we are interested in validating it experimentally. Informally, $H$ is falsifiable if it makes \textit{some} prediction that can (in principle, maybe not in practice) be checked unambiguously if it is right or wrong. This means that it should exist at least one experiment $\fE$ admitting at least one result which, if measured, can unambiguously confirm that $H$ is false, but not necessarily whether it is true. For instance, mechanical energy conservation for translation invariant systems is a prediction of Newtonian physics. Confirming it in one experiment does not confirm the truth of Newtonian physics, but (unambiguously) measuring a failure of it would mean that it is false.

By an abuse of notation, we will attribute a binary value to the variable $H$ which will stand for the truth value of the hypothesis. If $H=T$ the hypothesis is true, if $H=F$, it is false. If $H$ makes a falsifiable prediction, this means that we can separate all possible measurements of $\fE$ in two groups which we call $A$ and $B$. If $H$ predicts $B$, this means that if $H$ is true, $B$ is the only possible result of the measurement. Alternatively,a result $A$ cannot be obtained if $H$ is true. This can be written as the following condition on the probability of a measurement of $\fE$ given $H$ (the likelihood factor)
\begin{equation}
  \prob{\fE=A|H=T}=0,
  \label{equation:A|T}
\end{equation}
where we are again abusing the notation to attribute a binary value to $\fE$. By Bayes' rule, this directly implies that $\prob{H=T|\fE=A}=0$, or that if the result $A$ is obtained, the probability that the theory is true is zero. However, if the measurement falls on group $B$, it does not mean that the theory has been proved true. This is the subtle point that makes falsifiability a strong idea in science. Because one can never exhaust all possible measurements, there is always the possibility that, the next time, the theory will be proven wrong.

There is another condition that we require for a hypothesis to be falsifiable. If the result $A$ can never be obtained, then the experiment $\fE$ is meaningless as a way to falsify $H$. This leads to another condition on the likelihood
\begin{equation}
  \prob{\fE=A|H=F}\neq0.
  \label{equation:A|F}
\end{equation}

Let us call equations (\ref{equation:A|T}) and (\ref{equation:A|F}) respectively  the first and second falsifiability conditions. They are all that is required to say that the experiment $\fE$ falsifies $H$. If there is not such an experiment, $H$ is not falsifiable. The question now is whether we can say anything about the probability of $H$ being \textit{true} if we measure $B$. 

In order to save space in the expressions that follow, we will use a subindex to specify the random variables of the probability distributions and use as argument their values when convenient. For instance, the two falsifiability conditions become
\begin{equation}
  \probi{\fE|H}{A|T}=0, \qquad \probi{\fE|H}{A|F}\neq 0.
\end{equation}

Then, using Bayes' rule, we can write
\begin{equation}
  \probi{H|\fE}{T|B}=\frac{\probi{\fE|H}{B|T}\probi{H}{T}}{\probi{\fE|H}{B|F}\probi{H}{F}+\probi{\fE|H}{B|T}\probi{H}{T}}.
\end{equation}

Because $\probi{\fE|H}{A|h}+\probi{\fE|H}{B|h}=1$, $\forall h\in\chs{F,T}$, the first falsifiability condition implies $\probi{\fE|H}{B|T}=1$, while the second one gives $\probi{\fE|H}{B|F}<1$. We then can rewrite it as
\begin{equation}
  \probi{H|\fE}{T|B} = \frac{\probi{H}{T}}{\probi{\fE|H}{B|F}\prs{1-\probi{H}{T}}+ \probi{H}{T}}.
  \label{equation:PHE}
\end{equation}

Now, if we can guarantee that the above denominator is less than one, this would mean that $\probi{H|\fE}{T|B}>\probi{H}{T}$ and a measurement of $B$ would increase the probability that $H$ is true. This happens when
\begin{equation} 
  \probi{\fE|H}{B|F}\prs{1-\probi{H}{T}}+ \probi{H}{T} <1 \Rightarrow \probi{\fE|H}{B|F}<1,
\end{equation}
which we already derived as a consequence of the second condition. It is also possible to find the rate at which $\probi{H|\fE}{T|B}$ increases, the explicit calculation of which we leave to appendix \ref{appendix:AP}.

The result confirm the fact that, if a hypothesis is falsifiable by an experiment, by confirming its prediction on that experiment one gains confidence in the theory. The formal proof given above shows that Bayesian inference confirms our intuition in this case. Although $H$ can never be confirmed, we will say that $H$ is \textit{asymptotically provable} in this case. 

There are two points that need to be highlighted. The first is that, once a hypothesis has been falsified, it is ruled out for good. Bayes' rule implies that, if the prior for a result is zero, no amount of additional evidence can change it. The second concerns actual experiments. The above proof is based on the fact that the obtained measurements are trustworthy, which is rarely the case in practice. In a real situation, noise is usually present. This should be taken into consideration and, if done correctly, will change the equations above. This however will not be discussed further here. For our purposes, it is enough that the fundamental result stands in principle.

\section{Conditional Asymptotic Provability}

If the first condition on falsifiability is not met, the previous results are not valid and the hypothesis cannot be falsified. In particular, the following ratio
\begin{equation}
	\frac{\probi{H|\fE}{T|A}}{\probi{H|\fE}{F|A}} = \frac{\probi{\fE|H}{A|T}\probi{H}{T}}{\probi{\fE|H}{A|F}\probi{H}{F}}, 
\end{equation}
will now depend on the precise value of the factor $\phi\equiv\probi{\fE|H}{A|T}/\probi{\fE|H}{A|F}$. When the experiment could falsify the hypothesis, the precise value of the probability $\probi{\fE|H}{A|F}$ was not necessary, as long as it was not zero, because $\phi$ would be automatically zero. This is not true anymore. To know whether the above ratio increases or decreases with measurements, we need to know whether $\phi>1$ or $\phi<1$. However, unless we know all possible alternative hypotheses to $H$, this cannot be known. Therefore, if the hypothesis is not falsifiable by the experiment, the measurements will only be able to rank a set of competing hypotheses $h_i$, $i=1,2,...$, for which $\probi{\fE|H}{A|h_i}$ is known.

An alternative way to make possible attributing a numerical value to $\phi$ has been suggested by Dawid \cite{Dawid13,Dawid18}. Instead of providing a list of alternative hypotheses, he suggests to restrict the scope of the original hypothesis by analysing its truth for a certain domain which  he calls an \textit{empirical horizon}. The empirical horizon effectively restricts the possible experiments to the extent that $\phi$ can be calculated. 

For our purposes, it will be enough to assume that attributing a value to $\phi$ is possible in principle, but it might not be accessible to to us. This is because we are interested in a non-constructive proof of the possibility of falsification and its variations, but not on the specific problem of attributing numerical values for the confidence in the hypotheses. The latter is a problem that has no general solution as discussed by Dawid in the references cited above. Notice that impossibility of knowing the value $\phi$ does not preclude the asymptotic provability of falsifiable hypotheses, but it does forbid us of exactly quantifying the odds of it being true against being false and the same observations hold in this case too. 

Let us suppose now that there is another hypothesis $G$ which is falsifiable by an experiment $\fE$. Let us revert for a moment to the probability notation without subindices for convenience. If we can relate the truth of $G$ to that of $H$ by means of the conditional probability $\prob{H|G}$, then we would be able to write
\begin{equation}
  \begin{split}
    \prob{H|\fE} &= \sum_G \prob{H,G|\fE}\\
                         &= \sum_G \prob{H|G,\fE}\prob{G|\fE}\\
      					 &= \sum_G \prob{H|G}\prob{G|\fE},
  \end{split}
\end{equation}
where we assumed that $\fE$ has no direct influence on $H$ except through $G$. Back to subindex notation, let us define the $\probi{H|G}{T|T}=p$ and $\probi{H|G}{T|F}=q$ and write 
\begin{equation}
  \probi{H|\fE}{T|x} = p\probi{G|\fE}{T|x}+q\probi{G|\fE}{F|x}, \qquad x\in\chs{A,B}.
  \label{equation:pq}
\end{equation}

An additional requirement is that $p\neq q$, otherwise $H$ is independent of $G$. In order to make the arguments clear, let us connect the following reasoning to the case of the interior and exterior regions of a black hole. Previously, we used $O$ to denote the region outside the black hole (before the horizon) and $I$ to denote its interior (after the horizon) and we wrote $O\rightarrow I$ to symbolise that information can flow from $O$ to $I$ but not the opposite. The hypothesis we would like to assess, which however is not falsifiable, is the existence of the region $I$, i.e. $H=T$ if $E(I)=1$ and $H=F$ if $E(I)=0$. What we need then is to find a suitable hypothesis $G$ on which we can condition $H$.

There is a widespread belief that nature is not a patchwork, in the sense that observations seem to show that physical theories either remain the same or change smoothly when transitioning from one region of space to the other. It is possible that the universe is composed of different phases separated by domain walls \cite{Ellis04}, but even though there are underlying physical principles that are assumed to be valid in all regions. More radical hypotheses where domains have random physical laws \cite{Smolin92} are very controversial, especially because there is no way to assess them even probabilistically at the moment. 

Let us call the property that nature is not a patchwork of theories \textit{theoretical continuity} (TC) and use it as our conditioning hypothesis $G$. TC is falsifiable as it makes a precise prediction which can, in principle, be tested. Most cases of isolated regions in cosmology can be conditioned on this property. There is at present no indication that different patches of our universe sustain different physical laws, but measurements in principle could prove it to be wrong.  

If the experiment $\fE$ can falsify $G$, then (using our previous notation) measuring a value $A$ implies that $G$ is false and, therefore
\begin{equation}
  \probi{H|\fE}{T|A} = q.
\end{equation}

If $q=0$, then a measurement of $A$ also means that $H$ is false. In other words, $H$ can be falsified if the result is $A$. There is a subtlety though, $\probi{H|G}{T|F} = 0$ is equivalent to the logical statement $G=F\implies H=F$, which means that $H=T\implies G=T$. However, this last statement does not correspond to the situation we are trying to model. For our example, this would mean that the existence of region $I$ would necessarily imply that continuity is true everywhere in the universe, which does not need to be the case. In fact, if continuity is not true, one can easily have a situation in which $I$ exists.

If $q\neq0$ though, $H$ cannot be directly falsified by measuring $A$. Let us then consider the ratio
\begin{equation}
  \frac{\probi{H|\fE}{T|B}}{\probi{H|\fE}{F|B}}
    =\frac{p\probi{G|\fE}{T|B}+q\probi{G|\fE}{F|B}}{(1-p)\probi{G|\fE}{T|B}+(1-q)\probi{G|\fE}{F|B}},
\end{equation}
which we can write in terms of $\gamma\equiv\probi{G|\fE}{T|B}/\probi{G|\fE}{F|B}$ as
\begin{equation}
  \frac{\probi{H|\fE}{T|B}}{\probi{H|\fE}{F|B}} =\frac{p\gamma+q}{(1-p)\gamma+(1-q)}.
\end{equation}

From our previous results on falsifiability, we know that if $G$ is falsifiable, then the ratio $\gamma$ increases monotonically if we keep measuring only $B$. If this happens, the above ratio has three asymptotic limits. The first case is $p=0$, which leads to zero. This case, however, is equivalent to the one with $q=0$ if, instead of $G$, we condition all probabilities on the hypothesis $G'=\neg G$, the negation of $G$. This transformation also makes unnecessary to consider the case $q=1$ and we can assume $q\neq1$ as well. 

The second case is when $p\neq1$, which leads to $p/(1-p)$ and means that the hypothesis $H$ cannot be proved or disproved even asymptotically. Some probabilistic assessment can be made, but there is a fundamental limit to the uncertainty about $H$ which cannot be surpassed. Because there is still \textit{some} probabilistic assessment that can be done using Bayesian inference, we will say that in this case the hypothesis is \textit{Bayesian accessible}. 

The most interesting case is when $p=1$. Then, the ratio grows indefinitely with $\gamma$, which means that $H$ becomes \textit{conditionally} asymptotically provable. It is only in this case that the possibility of validating $H$ asymptotically given the falsifiable condition $G$ exists. For the case of the black hole, this would mean that the existence of $I$ would be asymptotically provable only in the case that TC implies it. In practice, that is what is usually done without the present formal justification.
 
\section{Conditioning Properties}

For the black hole case, we suggested that theoretical continuity is the property that should be used to condition the existence of the spacetime region $I$ such that $O$ can assess it rationally. It is also the condition that one would naturally use in the case of cosmological horizons caused by the expansion of the universe, the branching out of baby universes or Rindler horizons. For all this cases, we are usually reluctant to accept that the spacetime beyond the horizon has been destroyed by the phenomenon that created it. In the case of the Rindler horizon, in particular, the process is reversible and it seems unreasonable to assume that the spacetime was destroyed by acceleration and then re-created by deceleration. Rigorously, that remains a possibility though, as improbable as it might seem.

TC however would not be so straightforwardly justifiable in the case when pocket universes appearing independently as in eternal inflation and similar proposals. If the isolated regions haven't been in contact at any time in the past, TC alone is not an appealing property to attribute to physical theories in both regions. A more general property would be desirable.

Another possible candidate for a falsifiable property on which others could be conditioned would be \textit{mathematical consistency}. A mathematically rigorous definition of consistency is tricky \cite{Chang90}, but roughly speaking, one usually requires at least two features from it. The first is that one cannot prove at the same time a proposition and its negation to be true. Some care must be taken when constructing the propositions and how truth values are attributed to it though. The second, more appropriate to our beliefs about nature, is that if a prediction is derived by a theory using one method, the same result should be obtained by using any alternative methods allowed by that theory. As a certain result can always be reformulated as a yes/no question, the two definitions are equivalent.

Scientific models are constructed by using mathematical structures with well defined sets of axioms. Therefore, a mathematical inconsistency on the latter would result in an inconsistency of the former. The belief in the consistency of physical laws comes from the observational fact that it seems to hold in all natural phenomena ever studied. This belief is so strong that the apparent mathematical inconsistency between GR and quantum field theory under certain extreme conditions is one of the driving forces in the pursuit of a theory of quantum gravity \cite{Kiefer07}. It is worth noticing that, on philosophical grounds, the debate on whether consistency is necessary is still open \cite{Frisch14,daCosta14,Saatsi14}, but the great majority of scientists view inconsistencies as flaws in the models, although it would not prevent their use in practice if a better theory is lacking.

The modern theoretical position, again based on observations of nature up to this day, is that a mathematically inconsistent theory should be considered wrong from start, even though it might be successful in a limited way. However, the consistency criteria itself, although falsifiable and apparently highly probable, can never be \textit{proved} to be correct in an absolute sense, unless in the special case in which all possible experiments have already been done.

The list of possible properties does not end here. Symmetries and conservation laws are other possibilities. Notice that they also depend on the mathematical consistency of a theory, which creates a tree of dependencies. The best possible scenario would be to identify a minimal set of falsifiable properties, particularly for the case of fundamental theories. Recently, there has been a discussion about extra-empirical properties as guides to fundamental theories \cite{Pas19}. Falsifiability itself falls into this category together with naturalness, simplicity and elegance. Excluding falsifiability itself, whether or not a correct description of nature depends on these properties might make some of them falsifiable, while others not. An in depth analysis of their classification and how they are interrelated however is a complex subject and left for future work.

Conditioning as proposed here is closely related to one of Dawid's non-empirical confirmation arguments, the Meta-Inductive Argument \cite{Dawid18}. His meta-level hypotheses are equivalent to our conditioning ones, with the difference that we however only consider conditioning hypotheses which can be empirically assessed. 

\section{Discussion and Conclusions}
\label{section:Conc}

To summarise the work presented in here, we first pointed out that several theories of modern physics predict the existence of regions of spacetime which cannot exchange information with others. Although most of these theories are only tentative at the present, there is at least one well-tested falsifiable theory among them -- general relativity. If one region $M_1$ cannot receive information from another $M_2$, the existence of the latter cannot be falsified by any observer in the former. By assuming that a concept of absolute existence makes sense in nature, we arrived at what we called the \textit{isolated regions problem}: although $M_2$ exists, if its existence cannot be falsified by observers in $M_1$, strictly speaking $M_2$ cannot be scientifically assessed from $M_1$ and, ideally, should not be considered a scientific hypothesis in the latter. 

The solution we proposed for this conundrum is that probabilistic inference methods can extend the area of the hypothesis space which can be rationally assessed to include hypotheses which are, strictly speaking, not falsifiable. In order to do that, we introduced two sufficient conditions for a certain hypothesis to be falsifiable in probabilistic terms. We also prove that falsifiable hypotheses obeying these conditions can be asymptotically proved, i.e., each time their predictions are confirmed, the probability that they are true increases with a limit valued proved to be 1. We then suggested to use a falsifiable physical property $G$ to condition the existence of a hypothesis $H$ which is not falsifiable, deriving an equation for the change in the ration between the posteriors of $H$ being true or false. Although the formalism shows that $H$ cannot be falsifiable even conditionally, we could demonstrate that one can still make probabilistic statements about $H$. In particular, we show that if $\probi{H|G}{T|T}=1$, then $H$ is asymptotically provable conditioned on $G$. These developments have been guided by and applied to the case of spacetime regions.  

\begin{figure}
	\centering
	\includegraphics[width=7cm]{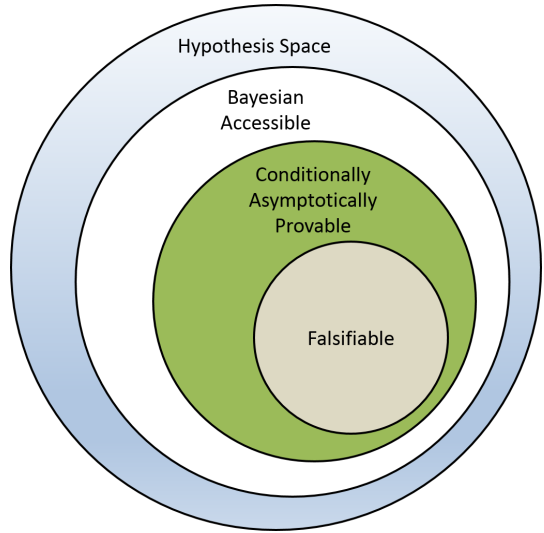}
	\caption{Hypothesis space. Each circle delimits a subset of hypotheses according to the possibility of evaluating their truth values. } 
	\label{figure:hspace}
\end{figure}

There are a series of consequences that can be derived from this proposal. The first is that Bayesian methods can extend the subset of the hypothesis space for which mathematically rigorous assessments can be carried out. The obtained results suggests that the space of all possible hypotheses can be divided in the regions depicted in fig. \ref{figure:hspace}.

The falsifiable subset of hypotheses is the one for which we can make the stronger statements. It contains the hypotheses which actually make testable predictions. Hypotheses which are conditionally asymptotic provable are not necessarily falsifiable, although they can be. Still, one can assign probabilities to their truth value (even if only in principle) which can asymptotically reach certainty under the conditions we specified. Even if one cannot asymptotically prove a hypotheses, we showed that one can still use probabilistic inference to assess them in a meaningful way in most cases. Such hypotheses, including those which fall outside the previous two regions, we call \textit{Bayesian accessible}. Whether or not there is any hypothesis outside the Bayesian accessible set remains an unanswered question which will not be addressed here.

The existence of isolated regions which are not falsifiable is not a demonstration that scientific reasoning based on the falsifiability criteria should be abandoned. It is rather an example of the limitations of the concept of falsifiability when applied to a \textit{plausible} physical model. It is important to notice that falsifiability is a condition of utmost importance to differentiate scientific knowledge from pseudo-science (or from superstition and religion). 

We want to stress that the results we got depend heavily on our assumed hypotheses. For instance, if absolute existence cannot be meaningfully defined, i.e., if there is no way to formulate a mathematical function $E$ that gives an absolute notion of existence, then the isolated world problem is ill-posed. One can view such a discussion as related that of \textit{reality}. This problem is particularly subtle in quantum theory. Discussions about whether there are levels of reality \cite{Busch10,Jaeger19} show that the question is still not settled. 

Another point that should be questioned is the existence of the conditional link between a falsifiable and a not falsifiable hypothesis. It seems that this link is simply equivalent to the assertion that the property should be valid everywhere and, therefore, is a falsifiable hypothesis. This is hardly a rigorous proof though and we leave the possibility open.

Finally, we have to leave very clear the fact that the statements about asymptotic provability are valid as limits. It is true that, in order to exactly quantify the probability of a hypothesis being true one would need the exact value of $\probi{\fE|H}{A|F}$, which as discussed before is only possible if one knows all alternative theories.


\section*{Acknowledgements}

I would like to thank Dr Juan Neirotti useful discussions.

\appendix

\section{Asymptotic Provability Limit}
\label{appendix:AP}

Consider equation (\ref{equation:PHE}). Let us us define, for convenience, $\probi{H}{T}=x_t$, $\probi{H|\fE}{T|B}=x_{t+1}$ and $\probi{\fE|H}{B|F}=\beta$. Then, the Bayesian update $x_t\to x_{t+1}$ is given by
\begin{equation}
x_{t+1} = \frac{x_t}{\beta(1-x_t)+x_t}.
\end{equation}

Suppose that the expression  
\begin{equation}
x_n = \frac{x_0}{\beta^k(1-x_0)+x_0},
\label{equation:xn}
\end{equation}
is true for some time $n$ with $k$ being an integer. Then
\begin{equation}
\begin{split}
x_{n+1} &= \frac{x_n}{\beta(1-x_n)+x_n}\\
&= \frac{x_0}{\beta^k(1-x_0)+x_0}\frac1{\beta\prs{1-\frac{x_0}{\beta^k(1-x_0)+x_0}}+\frac{x_0}{\beta^k(1-x_0)+x_0}}\\
&= \frac{x_0}{\beta^k(1-x_0)+x_0}\frac{\beta^k(1-x_0)+x_0}
{\beta\prs{\beta^k(1-x_0)+x_0-x_0}+x_0}\\
&=\frac{x_0}{\beta^{k+1}(1-x_0)+x_0}.
\end{split}
\end{equation}

But expression (\ref{equation:xn}) is indeed valid for $n=k=1$ and, therefore, this proves that
\begin{equation}
x_t = \frac{x_0}{\beta^t(1-x_0)+x_0},
\end{equation}
for any time $t$. Finally, the second falsifiability condition implies $\beta<1$, which then results in
\begin{equation}
\lim_{t\to\infty} x_t =1,
\end{equation}
in the case when only the result $B$ keeps being measured as long as $x_0\neq0$, which we assume true \textit{a priori}. This proves in a more direct way that the certainty about the truth of the hypothesis $H$ increases asymptotically to 1 if its predictions continue to be confirmed.


\bibliographystyle{spmpsci}       
\bibliography{worlds}   

\end{document}